\documentclass[11pt]{article}
\usepackage[margin=1in]{geometry}  
\usepackage{graphics, graphicx, color}
\usepackage{amssymb, amsfonts, amsthm, amsmath, mathrsfs}
\usepackage{multirow, booktabs}
\usepackage{subcaption}
\usepackage{algorithm}
\usepackage{algorithmic}
\usepackage[toc,page]{appendix}
\usepackage{authblk}
\usepackage{natbib}
\usepackage[colorlinks=true, linkcolor=blue, citecolor=blue]{hyperref}
\usepackage{url}
\usepackage{bm}

\newtheorem{proposition}{Proposition}

\theoremstyle{remark}

\theoremstyle{plain}
\newcommand{\diag}{\mbox{diag}}
\newcommand{\cov}{\mbox{cov}}

\newcommand\dd{\mathrm{d}}

\graphicspath{{Images/}}

\title{Energetic Variational Gaussian Process Regression for Computer Experiments}

\author[1]{Lulu Kang\thanks{lulukang@umass.edu}}
\affil[1]{Department of Mathematics and Statistics, University of Massachusetts, Amherst}
\author[2]{Yuanxing Cheng}
\affil[2]{Department of Applied Mathematics, Illinois Institute of Technology}
\author[3]{Yiwei Wang}
\affil[3]{Department of Mathematics, University of California, Riverside}
\author[2]{Chun Liu}

\begin{document}

\maketitle

\begin{abstract}
The Gaussian process (GP) regression model is a widely employed surrogate modeling technique for computer experiments, offering precise predictions and statistical inference for the computer simulators that generate experimental data. 
Estimation and inference for GP can be performed in both frequentist and Bayesian frameworks. 
In this chapter, we construct the GP model through variational inference, particularly employing the recently introduced energetic variational inference method by \cite{wang2021particle}. 
Adhering to the GP model assumptions, we derive posterior distributions for its parameters.  
The energetic variational inference approach bridges the Bayesian sampling and optimization and enables approximation of the posterior distributions and identification of the posterior mode. 
By incorporating a normal prior on the mean component of the GP model, we also apply shrinkage estimation to the parameters, facilitating mean function variable selection. To showcase the effectiveness of our proposed GP model, we present results from three benchmark examples.
\end{abstract}

\section{Introduction}

Uncertainty Quantification \citep{ghanem2017handbook} is a highly interdisciplinary research domain that involves mathematics, statistics, optimization, advanced computing technology, and various science and engineering disciplines. 
It provides a computational framework for quantifying input and response uncertainties and making model-based predictions and their inferences for complex science or engineering systems/processes. 
One fundamental research problem in uncertainty quantification is to analyze computer experimental data and build a surrogate model for the computer simulation model.
Gaussian process (GP) regression model, sometimes known as ``kriging'', has been widely used for this purpose since the seminal paper by \cite{sacks_design_1989}. 

In this book chapter, we plan to re-introduce Gaussian process models under the Bayesian framework. 
Among the various model assumptions, we choose the one from \cite{santner2003design}. 
It has a mean function that can be helpful in model interpretation and an anisotropic covariance function that provides more flexibility to improve prediction accuracy. 
Different from most existing works on Bayesian GP models, we provide an alternative computational tool based on \emph{variational inference} to replace the common Markov Chain Monte Carlo (MCMC) sampling method.
Specifically, we propose using a particle-based energetic variational inference approach (EVI) developed in \cite{wang2021particle} to generate samples and approximate the posterior distribution. 
For short, we name the method EVI-GP. 
Depending on the number of particles, EVI-GP can be used to compute the maximum a posteriori (MAP) estimate of the parameters or obtain a large number of samples to approximate the posterior distribution of the parameters. 
The estimations and inference of parameters and predictions can be obtained from the MAP estimate or the posterior samples. 
Thanks to the conjugate prior of the regression coefficients of the mean function, $l_2$-regularization is adopted to achieve model sparsity for the mean function of the GP. 

The rest of the chapter is organized as follows. 
In Section \ref{sec:GP}, we review the Gaussian process model, including its assumption and prior distributions, and derive the posterior and posterior predictive distributions. 
In Section \ref{sec:EVIGP}, the preliminary background on variational inference and the particle energetic variational inference methods are briefly reviewed. 
The EVI-GP method is also summarized at the end of this section. 
Section 4 shows three main simulation examples in which different versions of EVI-GP are compared with other existing methods. 
The chapter concludes in Section 5. 

\section{Gaussian Process Regression: Bayesian Approach}\label{sec:GP}

\subsection{Gaussian Process Assumption}\label{subsec:assump}
Denote $\{\bm x_i, y_i\}_{i=1}^n$ as the $n$ pairs of input and output data from a certain computer experiment, and $\bm x_i\in \Omega \subseteq \mathbb{R}^d$ are the $i$th experimental input values and $y_i\in \mathbb{R}$ is the corresponding output. 
In this chapter, we only consider the case of univariate response, but the proposed EVI-GP can be applied to the multi-response GP model which involves the cross-covariance between responses \citep{cressie2015statistics}. 

Gaussian process regression is built on the following model assumption of the response, 
\begin{equation}\label{eq:model}
y_i=\bm g(\bm x_i)^\top \bm \beta+ Z(\bm x_i)+\epsilon_i, \quad i=1,\ldots, n,
\end{equation}
where $\bm g(\bm x)$ is a $p$-dim vector of user-specified basis functions and $\bm \beta$ is the $p$-dim vector of linear coefficients corresponding the basis functions. 
Usually, $\bm g(\bm x)$ contains the polynomial basis functions of $\bm x$ up to a certain order. 
For example, if $\bm x \in \mathbb{R}^2$, $\bm g(\bm x)=[1, x_1, x_2, x_1^2, x_1x_2, x_2^2]$ contains the intercept, the linear, the quadratic and the interaction effects of the two input variables. 
The random noise $\epsilon_i$'s are independently and identically distributed following $\mathcal{N}(0, \sigma^2)$. 
They are also independent of the other stochastic components of \eqref{eq:model}. 
We assume the GP prior on the stochastic function $Z(\bm x)$, which is denoted as $Z(\cdot)\sim GP(0, \tau^2 K)$, i.e., $\mathbb{E}[Z(\bm x)]=0$ and the covariance function
\[
\cov[Z(\bm x_1), Z(\bm x_2)]=\tau^2 K(\bm x_1, \bm x_2;\bm \omega).
\]
In most applications of computer experiments, we use the stationary assumption of $Z(\bm x)$, and thus the variance $\tau^2$ is a constant. 
The function $K(\cdot, \cdot;\bm \omega): \Omega \times \Omega \mapsto \mathbb{R}_{+}$ is the correlation of the stochastic process with hyperparameters $\bm \omega$. 
For it to be valid, $K(\cdot,\cdot;\bm \omega)$ must be a symmetric positive definite kernel function. 
Gaussian kernel function is one of the most commonly used kernel functions and its definition is 
\[
K(\bm x_1, \bm x_2; \bm \omega)=\exp\left\{-\sum_{j=1}^d \omega_j(x_{1j}-x_{2j})^2\right\},
\]
with $\bm \omega \in \mathbb{R}^d$ and $\bm \omega\geq 0$. 
Although we only demonstrated the examples using the Gaussian kernel, the EVI-GP method can be applied in the same way for other kernel functions. 

In terms of response $Y(\bm x)$, it follows a Gaussian Process with the following mean and covariance, 
\begin{align}\label{eq:Ymodel}
\mathbb{E}[Y(\bm x)]&=\bm g(\bm x)^\top \bm \beta, \quad \forall \bm x \in \Omega\\
\cov[Y(\bm x_1), Y(\bm x_2)]&=\tau^2 K(\bm x_1, \bm x_2;\bm \omega) + \sigma^2\delta(\bm x_1, \bm x_2), \quad \forall \bm x_1, \bm x_2\in \Omega,\\\
&=\tau^2\left[ K(\bm x_1, \bm x_2;\bm \omega) + \eta\delta(\bm x_1, \bm x_2)\right], 
\end{align}
where $\delta(\bm x_1,\bm x_2)=1$ if $\bm x_1=\bm x_2$ and 0 otherwise, and $\eta=\sigma^2/\tau^2$. 
So $\eta$ is interpreted as the noise-to-signal ratio. 
For deterministic computer experiments, the noise component is not part of the model, i.e., $\sigma^2=0$. 
However, a nugget effect, which is a small $\eta$ value, is usually included in the covariance function to avoid the singularity of the covariance matrix \citep{peng2014on}. 
The unknown parameter values of the GP model are $\bm \theta=(\bm \beta, \bm \omega, \tau^2, \eta)$. 
We are going to show how to obtain the estimation and inference of the parameters using the Bayesian framework. 

\subsection{GP under Bayesian Framework}\label{sub:bayesianGP}
We assume the following prior distributions for the parameters $\bm \theta=(\bm \beta, \bm \omega, \tau^2, \eta)$, 
\begin{align}\label{eq:beta-prior}
& \bm \beta \sim \mathcal{MVN}_p({\bf 0}, \nu^2 \bm R)\\\label{eq:omega-prior}
& \omega_i \overset{\text{i.i.d.}}{\sim} \text{Gamma}(a_{\omega}, b_{\omega}), \text{ for } i=1,\ldots, d \\\label{eq:tau-prior}
& \tau^2 \sim \text{Inverse-}\chi^2(df_{\tau^2}), \\\label{eq:eta-prior}
& \eta \sim \text{Gamma}(a_{\eta}, b_{\eta}).
\end{align}
These distribution families are commonly used in the literature such as \cite{gramacy2008bayesian,gramacy2015local}. 
However, the choice of parameters of the prior distributions should require fine-tuning using testing data or cross-validation procedures. 
In some literature, parameters $\bm \omega$, $\tau^2$, and $\eta$ are considered to be \emph{hyperparameters}. 
The conditional posterior distribution of $\bm \beta$ given data and $(\bm \omega, \tau^2, \eta)$ is a multivariate normal distribution, which is shown later. 

Next, we derive the posterior distributions and some conditional posterior distributions. 
Based on the data, the sampling distribution is 
\begin{equation}
\bm y_n|\bm \theta \sim \mathcal{MVN}_n\left(\bm G\bm \beta, \tau^2(\bm K_n+\eta \bm I_n)\right),
\end{equation}
where $\bm y_n$ is the vector of $y_i$'s and $\bm G$ is a matrix of row vectors $\bm g(\bm x_i)^\top$'s. 
The matrix $\bm K_n$ is the $n\times n$ kernel matrix with entries $K_n[i,j]=K(\bm x_i,\bm x_j)$ and is a symmetric and positive definite matrix, and $\bm I_n$ is an $n\times n$ identity matrix. 
The density function of $\bm y_n|\bm \theta$ is 
\begin{equation}
p(\bm y_n|\bm \theta) \propto (\tau^2)^{-\frac{n}{2}}\det(\bm K_n+\eta\bm I_n)^{-1/2}\exp\left(-\frac{1}{2\tau^2}(\bm y_n-\bm G\bm \beta)^\top (\bm K_n+\eta \bm I_n)^{-1} (\bm y_n-\bm G\bm \beta)\right).
\end{equation}
Following Bayes' Theorem, the joint posterior distribution of all parameters is 
\begin{align*}
p(\bm \theta|\bm y_n)&\propto p(\bm \theta) p(\bm y_n|\bm \theta)\\
&\propto p(\bm \beta)\left(\prod_{j=1}^d p(\omega_i)\right) p(\tau^2) p(\eta) p(\bm y_n|\bm \theta).
\end{align*}

The conditional posterior distribution of $\bm \beta$ can be easily obtained through conjugacy. 
It is also straightforward to obtain the posterior distribution $p(\bm \omega, \tau^2,\eta|\bm y_n)$ as shown in the appendix. 
The results are summarized in Proposition \ref{prop:post}. 
\begin{proposition}\label{prop:post}
Using the prior distribution of $\bm \theta=(\bm \beta, \bm \omega, \tau^2, \eta)$ in \eqref{eq:beta-prior}-\eqref{eq:eta-prior}, the conditional posterior distribution of $\bm \beta$ is 
\[
\bm \beta | \bm y_n, \bm \omega, \tau^2,\eta \sim \mathcal{MVN}_p\left(\hat{\bm \beta}_n, \bm \Sigma_{\bm \beta|n}\right),
\]
where 
\begin{align}\label{eq:beta-cov}
\bm \Sigma_{\bm \beta|n}&=\left[\frac{1}{\tau^2}\bm G^\top (\bm K_n+\eta \bm I_n)^{-1}\bm G+\frac{1}{\nu^2} \bm R^{-1}\right]^{-1},\\\label{eq:beta-m}
\hat{\bm \beta}_n&=\tau^{-2}\bm \Sigma_{\bm \beta|n}\left[\bm G^\top (\bm K_n+\eta \bm I_n)^{-1}\right]\bm y_n.
\end{align}
The marginal posterior distribution of $(\bm \omega, \tau^2, \eta)$ is 
\begin{align}\nonumber
p(\bm \omega, \tau^2,\eta|\bm y_n) \propto & \det(\bm \Sigma_{\bm \beta|n})^{1/2}\exp\left[-\frac{1}{2}\hat{\bm \beta}_n^\top \bm \Sigma_{\bm \beta|n}^{-1}\hat{\bm \beta}_n-\frac{1}{2\tau^2}\bm y_n^\top (\bm K_n+\eta \bm I_n)^{-1}\bm y_n\right]\\\label{eq:post-other}
&\times (\tau^2)^{-n/2}\det(\bm K_n+\eta \bm I_n)^{-1/2}p(\tau^2)p(\bm \omega)p(\eta). 
\end{align}
\end{proposition}

The posterior distribution of $\tau^2$ can be significantly simplified if a non-informative prior is used for $\bm \beta$, as described in Proposition \ref{prop:post2}. 
\begin{proposition}\label{prop:post2}
If using a non-informative prior distribution for $\bm \beta$, i.e., $p(\bm \beta)\propto 1$, and the prior distributions \eqref{eq:omega-prior}-\eqref{eq:eta-prior} for the remaining parameters, the conditional posterior distribution of $\bm \beta$ is still $\mathcal{MVN}_p\left(\hat{\bm \beta}_n, \bm \Sigma_{\bm \beta|n}\right)$, but the covariance and mean are 
\begin{align*}
\bm \Sigma_{\bm \beta|n}&=\tau^2\left[\bm G^\top (\bm K_n+\eta \bm I_n)^{-1} \bm G\right]^{-1},\\
\hat{\bm \beta}_n &= \left[\bm G^\top (\bm K_n+\eta \bm I_n)^{-1} \bm G\right]^{-1}\left[\bm G^\top (\bm K_n+\eta \bm I_n)^{-1}\right]\bm y_n.
\end{align*}
The conditional posterior distribution for $\tau^2$ is 
\begin{equation}\label{eq:tau-post2}
\tau^2 |\bm \omega, \eta, \bm y_n \sim \text{Scaled Inverse-}\chi^2(df_{\tau^2}+n-p, \hat{\tau}^2), 
\end{equation}
where 
\begin{align*}
\hat{\tau}^2&=\frac{1+s_n^2}{df_{\tau^2}+n-p},\\
s^2_n&=\tau^{-2} \hat{\bm \beta}_n^\top \bm \Sigma_{\bm \beta | n}^{-1}\hat{\bm \beta}_n+\bm y_n^\top (\bm K_n+\eta \bm I_n)^{-1}\bm y_n.
\end{align*}
The marginal posterior of $(\bm \omega, \eta)$ is 
\begin{align}\label{eq:post2-other}
p(\bm \omega,\eta |\bm y_n) \propto (\hat{\tau}^2)^{-\frac{1}{2}(df_{\tau^2}+n-p)}\det(\bm G^\top (\bm K_n+\eta \bm I_n)^{-1}\bm G)^{-1/2}\det(\bm K_n+\eta \bm I_n)^{-1/2}p(\bm \omega) p(\eta).
\end{align}
\end{proposition}

If we use non-informative prior distributions for all the parameters, i.e., $p(\bm \beta)\propto 1$, $p(\omega_i)\overset{\text{i.i.d.}}{\sim}\text{Uniform}[a_{\omega}, b_{\omega}]$ for $i=1,\ldots, d$, $p(\tau^2)\propto \tau^{-2}$, and $p(\eta)\propto \text{Uniform}[a_{\eta}, b_{\eta}]$, the Bayesian framework is equivalent to the empirical Bayesian or maximum likelihood estimation method. 
The GP regression model estimated via this \emph{frequentist} approach is common in both methodology research and application \citep{santner2003design,fang2005design,gramacy2020surrogates}. 
In this chapter, we consider the empirical Bayesian estimation as a special case of the Bayesian GP model. 
The choice between the two different types of prior distributions for $\bm \beta$, informative or non-informative, is subject to the dimension of the input variables, the assumption on basis functions $\bm g(\bm x)$, the goal of GP modeling (accurate prediction v.s. interpretation), and sometimes the application of the computer experiment.
Both types have their unique advantages and shortcomings. 
The non-informative prior distribution for $\bm \beta$ reduces the computation involved in the posterior sampling for $\tau^2$, but we would lose the $l_2$ regularization effect on $\bm \beta$ brought by the informative prior $\bm \beta$.

One issue with the informative prior distribution is to choose its parameters, i.e., the constant variance $\nu^2$ and the correlation matrix $\bm R$. 
Here we recommend using a cross-validation procedure to select $\nu^2$. 
If the mean function $\bm g(\bm x)^\top \bm \beta$ is a polynomial function of the input variables, we specify the matrix $\bm R$ to be a diagonal matrix $\bm R=\diag\{1, r, \ldots, r, r^2, \ldots, r^2, \ldots\}$, where $r\in (0,1)$ is a user-specified parameter.
The power index of $r$ is the same as the order of the corresponding polynomial term.
For example, if $\bm g(\bm x)^\top \bm \beta$ with $\bm x \in \mathbb{R}^2$ is a full quadratic model and contains the terms $\bm g(\bm x)=[1,x_1,x_2,x_1^2,x_2^2,x_1x_2]^\top$, the corresponding prior correlation matrix should be specified as $\bm R=\diag\{1,r,r,r^2,r^2,r^2\}$.
In this way, the prior variance of the effect decreases exponentially as the order of effect increases, following the \emph{effect hierarchy principle} \citep{hamada1992analysis, wu2021experiments}. 
It states that lower-order effects are more important than higher-order effects, and the effects of the same order are equally important.
The hierarchy ordering principle can reduce the size of the model and avoid including higher-order and less significant model terms.
Such prior distribution was firstly proposed by \cite{joseph2006bayesian}, and later used in \cite{kang2009bayesian, ai2009bayesian, kang2018bayesian, kang2019bayesian,kang2021bayesian,kang2023bayesian}.



\begin{proposition}\label{prop:pred}
Given the parameters $(\bm \omega, \tau^2, \eta)$, the posterior predictive distribution of $y(\bm x)$ at any query point $\bm x$ is the following normal distribution. 
\begin{equation}\label{eq:post-pred}
y(\bm x)|\bm y_n, \bm \omega, \tau^2, \eta \sim \mathcal{N}(\hat{\mu}(\bm x), \sigma^2_n(\bm x)), 
\end{equation}
where 
\begin{align}\label{eq:pred-mean}
\hat{\mu}(\bm x)&=\bm g(\bm x)^\top \hat{\bm \beta}_n+K(\bm x, \mathcal{X}_n) (\bm K_n+\eta \bm I_n)^{-1}(\bm y_n-\bm G\hat{\bm \beta}_n),\\
\label{eq:post-var}
\sigma_n^2(\bm x)&=\tau^2\left\{1-K(\bm x, \mathcal{X}_n)(\bm K_n+\eta \bm I_n)^{-1} K(\mathcal{X}_n, \bm x)+\bm c(\bm x)^\top \left[\bm G^\top (\bm K_n+\eta \bm I_n)^{-1}\bm G \right]^{-1}\bm c(\bm x)\right\}, 
\end{align}
where 
\begin{align*}
\bm c(\bm x)&=\bm g(\bm x)-\bm G^\top (\bm K_n+\eta \bm I_n)^{-1} K(\mathcal{X}_n, \bm x), \\
K(\bm x, \mathcal{X}_n)&=K(\mathcal{X}_n, \bm x)^\top=[K(\bm x, \bm x_1), \ldots, K(\bm x, \bm x_n)].
\end{align*}
For non-informative prior, the posterior predictive distribution $y(\bm x)|\bm y_n, \bm \omega, \eta$ is the same except $\tau^2$ is replaced by $\hat{\tau}^2$. 
\end{proposition}
A detailed proof can be found in \cite{santner2003design} and \cite{rasmussen_gaussian_2006}.

Thanks to the Gaussian process assumption and the conditional conjugate prior distributions for $\bm \beta$ and $\tau^2$, Proposition \ref{prop:post}, \ref{prop:post2}, and \ref{prop:pred} give the explicit and easy to generate conditional posterior distribution of $\bm \beta$ (and $\tau^2$) and posterior predictive distribution. 
Therefore, how to generate samples from $p(\bm \omega, \tau^2,\eta|\bm y_n)$ in \eqref{eq:post-other} or $p(\bm \omega, \eta|\bm y_n)$ in \eqref{eq:post2-other} is the bottleneck of the computation for GP models. 
Since $p(\bm \omega, \tau^2,\eta|\bm y_n)$ and $p(\bm \omega, \eta|\bm y_n)$ are not from any known distribution families, Metropolis-Hastings (MH) algorithm \citep{robert1999monte, gelman2014bayesian}, Hamiltonian Monte Carlo (HMC) \citep{Neal1996}, or Metropolis-adjusted Langevin algorithm (MALA) can be used to for sampling \citep{roberts1998optimal}.
In this chapter, we introduce readers to an alternative computational tool, namely, a variational inference approach to approximate the posterior distribution $p(\bm \omega,\tau^2,\eta|\bm y_n)$ or $p(\bm \omega, \eta|\bm y_n)$. 
More specifically, we plan to use energetic variational inference, a particle method, to generate posterior samples. 

\section{Energetic Variational Inference Gaussian Process}\label{sec:EVIGP}

Variational inference-based GP models have been explored in prior works such as \cite{tran2016variational}, \cite{cheng2017variational}, and \cite{wynne2022variational}.
Despite sharing the variational inference idea, the proposed Energetic Variational Inference (EVI) GP differs significantly from these existing methods regarding the specific variational techniques employed. 
\cite{tran2016variational} utilized auto-encoding, while \cite{cheng2017variational} and \cite{wynne2022variational} employed the mean-field variational inference \citep{blei2017variational}. 

The EVI approach presented in this chapter is a newly introduced particle-based method. 
It offers simplicity in implementation across diverse applications without the need for training any neural networks. 
Notably, EVI establishes a connection between the MAP procedure and posterior sampling through a user-specified number of particles. 
In contrast to the complexity of auto-encoding variational methods, the particle-based approach is much simpler, devoid of any network intricacies. 
Moreover, in comparison to mean-field methods, both particle-based and MAP-based approaches (auto-encoding falls into the MAP-based category) can exhibit enhanced accuracy, as they do not impose any parametric assumptions on a feasible family of distributions in the optimization to solve the variational problem. 

This section provides a brief introduction to the Energetic Variational Inference (EVI) approach. 
Subsequently, we adapt this method for the estimation and prediction of the GP regression model.

\subsection{Preliminary: Energetic Variational Inference}\label{subsec:EVI}

Motivated by the energetic variational approaches for modeling the dynamics of non-equilibrium thermodynamical systems \citep{Giga2017}, the energetic variational inference framework uses a continuous energy-dissipation law to specify the dynamics of minimizing the objective function in machine learning problems.
Under the EVI framework, a practical algorithm can be obtained by introducing a suitable discretization to the continuous energy-dissipation law.
This idea was introduced and applied to variational inference by \cite{wang2021particle}.
It can also be applied to other machine learning problems similar to \cite{trillos2018bayesian} and \cite{weinan2020machine}.

We first introduce the EVI using the continuous formulation.
Let $\bm \phi_t$ be the dynamic flow map $\bm \phi_t: \mathbb{R}^d \rightarrow \mathbb{R}^d$ at time $t$ that continuously transforms the $d$-dimensional distribution from an initial distribution toward the target one and we require the map $\bm \phi_t$ to be smooth and one-to-one.
The functional $\mathcal{F}(\bm \phi_t)$ is a user-specified divergence or other machine learning objective functional, such as the KL-divergence in \cite{wang2021particle}.
Taking the analogy of a thermodynamics system, $\mathcal{F}(\bm \phi_t)$ is the Helmholtz free energy.
Following the First and Second Law of thermodynamics \citep{Giga2017} (kinetic energy is set to be zero)
\begin{equation}\label{eq:energydiss}
\frac{\dd}{\dd t} \mathcal{F}(\bm \phi_t) = - \triangle(\bm \phi_t, \dot{\bm \phi}_t),
\end{equation}
where $\triangle(\bm \phi_t, \dot{\bm \phi}_t)$ is a user-specified functional representing the rate of energy dissipation, and $\dot{\bm \phi}_t$ is the derivative of $\bm \phi_t$ with time $t$.
So $\dot{\bm \phi}_t$ can be interpreted as the ``velocity'' of the transformation.
Each variational formulation gives a natural path of decreasing the objective functional $\mathcal{F}(\bm \phi_t)$ toward an equilibrium.

The dissipation functional should satisfy $\triangle(\bm \phi_t, \dot{\bm \phi}_t)\geq 0$ so that $\mathcal{F}(\bm \phi_t)$ decreases with time.
As discussed in \cite{wang2021particle}, there are many ways to specify $\triangle(\bm \phi_t, \dot{\bm \phi}_t)$ and the simplest among them is a quadratic functional in terms of $\dot{\bm \phi}_t$,
\[
\triangle(\bm \phi_t,\dot{\bm \phi}_t)=\int_{\Omega_t}\rho_{[\bm \phi_t]}\|\dot{\bm \phi}_t\|_2^2 \dd \bm x,
\]
where $\rho_{[\bm \phi_t]}$ denotes the pdf of the current distribution which is the initial distribution transformed by $\bm \phi_t$, $\Omega_t$ is the current support, and $\| {\bm a} \|_2 = {\bm a}^\top {\bm a}$ for $\forall {\bm a} \in \mathbb{R}^d$.
This simple quadratic functional is appealing since it has a simple derivative, i.e.,
\begin{equation*}
   \frac{\delta \triangle(\bm \phi_t,\dot{\bm \phi}_t)}{\delta \dot{\bm \phi}_t}= 2\rho_{[\bm \phi_t]} \dot{\bm \phi}_t,
\end{equation*}
where $\delta$ is the variation operator, i.e., functional derivative.

With the specified energy-dissipation law \eqref{eq:energydiss}, the energy variational approach derives the dynamics of the systems through two variational procedures, the Least Action Principle (LAP) and the Maximum Dissipation Principle (MDP), which leads to
\begin{equation*}
\frac{\delta \frac{1}{2}\triangle}{\delta \dot{\bm \phi}_t} = - \frac{\delta \mathcal{F}}{\delta \bm \phi_t}.
\end{equation*}
The approach is motivated by the seminal works of Raleigh \citep{strutt1871some} and Onsager \citep{onsager1931reciprocal,onsager1931reciprocal2}.
Using the quadratic $\mathcal{D}(\bm \phi_t, \dot{\bm \phi}_t)$, the dynamics of decreasing $\mathcal{F}$ is
\begin{equation}\label{eq:EVIv1}
\rho_{[\bm \phi_t]} \dot{\bm \phi}_t = - \frac{\delta \mathcal{F}}{\delta \bm \phi_t}.
\end{equation}

In general, this continuous formulation is difficult to solve, since the manifold of $\bm \phi_t$ is of infinite dimension.
Naturally, there are different approaches to approximate an infinite-dimensional manifold by a finite-dimensional manifold.
One such approach, as used in \cite{wang2021particle}, is to use \emph{particles} (or \emph{samples}) to approximate the $\rho_{[\bm \phi_t]}$ in \eqref{eq:energydiss} with kernel regularization, before any variational steps.
It leads to a discrete version of the energy-dissipation law, i.e.,
\begin{equation}\label{eq:EVIv2}
\frac{\dd}{\dd t} \mathcal{F}_h(\{\bm x_i(t)\}_{i=1}^N)=-\triangle_h(\{\bm x_i(t)\}_{i=1}^N, \{\dot{\bm x}_i(t)\}_{i=1}^N).
\end{equation}
Here $\{\bm x(t)\}_{i=1}^N$ is the locations of $N$ particles at time $t$ and $\dot{\bm x}_i(t)$ is the derivative of $\bm x_i$ with $t$, and thus is the velocity of the $i$th particle as it moves toward the target distribution.
The subscript $h$ of $\mathcal{F}$ and $\triangle$ denotes the bandwidth parameter of the kernel function used in the kernelization operation.
Applying the variational steps to \eqref{eq:EVIv2}, we obtain the dynamics of decreasing $\mathcal{F}$ at the particle level,
\begin{equation}\label{eq:EVIv3}
\frac{\delta \frac{1}{2}\triangle_h}{\delta \dot{\bm x}_i(t)} = - \frac{\delta \mathcal{F}_h}{\delta \bm x_i}, \quad \text{for }i=1, \ldots, N.
\end{equation}
This leads to an ODE system of $\{\bm x_i(t)\}_{i=1}^N$ and can be solved by different numerical schemes.
The solution is the particles approximating the target distribution.

Due to limited space, we can only briefly review the EVI framework and explain it intuitively.
Readers can find the rigorous and concrete explanation in \cite{wang2021particle}.
It also suggested many different ways to specify the energy dissipation, the ways to approximate the continuous formulation, and different ways to solve the ODE system.

\subsection{EVI-GP}

In this chapter, we use the KL-divergence as the energy functional as demonstrated in \cite{wang2021particle},
\begin{equation}
D_{\text{KL}}(\rho||\rho^*)=\int_{\Omega}\rho(\bm x) \log\left(\frac{\rho(\bm x)}{\rho^*(\bm x)}\right)\dd \bm x,
\end{equation}
where $\rho^*(\bm x)$ is the density function of the target distribution with support region $\Omega$ and $\rho(\bm x)$ is to approximate $\rho^*(\bm x)$. 
For EVI-GP, $\rho^*$ is the posterior distribution of $p(\bm \omega, \tau^2, \eta|\bm y_n)$ or $p(\bm \omega, \tau^2|\bm y_n)$. 

Using the KL-divergence, the divergence functional at time $t$ is 
\[\mathcal{F}(\bm \phi_t) = \int \left(\rho_{[\bm \phi_t]}(\bm x)\log\rho_{[\bm \phi_t]}(\bm x)+\rho_{[\bm \phi_t]}(\bm x) V(\bm x) \right)\dd\bm x,\]
where $V(\bm x)=\log \rho^*(\bm x)$, which is known up to a scaling constant. 
The discrete version of the energy becomes 
\begin{equation}\label{eq:F_h}
\mathcal{F}_h\left(\{\bm x_i\}_{i=1}^N\right) = \frac{1}{N} \sum_{i = 1}^N \left( \ln \left( \frac{1}{N}\sum_{j = 1}^N K_h(\bm x_i, \bm x_j) \right) + V(\bm x_i) \right),
\end{equation}
and the discrete dissipation is
\begin{equation}\label{eq:D_h}
-2\mathcal{D}_h\left(\{\bm x_i\}_{i=1}^N\right)=- \frac{1}{N} \sum_{i=1}^N |\dot{\bm x}_i(t))|^2.
\end{equation}
Applying variational step to \eqref{eq:EVIv2}, we obtain \eqref{eq:EVIv3} which is equivalent to the following nonlinear ODE system: 
\begin{equation}\label{ODE_blob}
\dot{\bm x}_i(t) =  - \left( \frac{ \sum_{j=1}^N \nabla_{\bm x_i} K_h(\bm x_i, \bm x_j)}{\sum_{j = 1}^N K_h(\bm x_i, \bm x_j) } + \sum_{k=1}^N \frac{\nabla_{\bm x_i} K_h(\bm x_k, \bm x_i)}{\sum_{j=1}^N K_h(\bm x_k, \bm x_j)}+ \nabla_{\bm x_i} V(\bm x_i) \right), \\
\end{equation}
for $i=1,\ldots, N$.
The iterative update of $N$ particles ${\bm x_i}_{i=1}^N$ involves solving the nonlinear ODE system \eqref{ODE_blob} via optimization problem \eqref{Min_problem} at the $m$-th iteration step
\begin{equation}\label{Min_problem}
\{\bm x^{m+1}_i \}_{i=1}^N = \text{argmin}_{\{\bm x_i\}_{i=1}^N} J_m(\{ \bm x_i \}_{i=1}^N),
\end{equation} 
where  
\begin{equation}\label{eq:Jn}
J_m( \{ \bm x_i \}_{i=1}^N):= \frac{1}{2 \tau} \sum_{i=1}^N ||\bm x_i - \bm x_i^m||^2 / N + \mathcal{F}_h( \{ \bm x_i \}_{i=1}^N).
\end{equation}
\cite{wang2021particle} emphasized the advantages of using the Implicit-Euler solver for enhanced numerical stability in this process.
We summarize the algorithm of using the implicit Euler scheme to solve the ODE system \eqref{ODE_blob} into Algorithm \ref{alg:EVI-Im}. 
Here MaxIter is the maximum number of iterations of the outer loop.
The optimization problem in the inner loop is solved by L-BFGS in our implementation.  
\begin{algorithm}[ht]
\caption{EVI with Implicit Euler Scheme (EVI-Im)}
\label{alg:EVI-Im}
\begin{algorithmic}
 \STATE {\bfseries Input:} The target distribution $\rho^{*}(\bm x)$ and a set of initial particles $\{\bm x_i^0\}_{i=1}^N$ drawn from a prior $\rho_0(\bm x)$.
 \STATE {\bfseries Output:} A set of particles $\{\bm x_i^{*}\}_{i=1}^N$ approximating $\rho^*$.
 \FOR{$m=0$ {\bfseries to} $\textrm{MaxIter}$}
 \STATE Solve $\{\bm x^{m+1}_i \}_{i=1}^N = \text{argmin}_{\{\bm x_i\}_{i=1}^N} J_m(\{ \bm x_i \}_{i=1}^N)$. 
 \STATE Update $\{ \bm x_i^{m}\}_{i=1}^N$ by $\{ \bm x_i^{m+1} \}_{i=1}^N$. 
 \ENDFOR
\end{algorithmic}
\end{algorithm}

We propose two adaptations of the EVI-Im algorithm for GP model estimation and prediction. 
The first approach involves generating $N$ particles using Algorithm \ref{alg:EVI-Im} to approximate the posterior $p(\bm \omega, \tau^2, \eta|\bm y_n)$ when an informative normal prior distribution is adopted for $\bm \beta$, or $p(\bm \omega, \tau^2|\bm y_n)$ when a non-informative prior distribution is used for $\bm \beta$. 
These $N$ particles serve as posterior samples. 
Conditional on their values, we can generate samples for $\bm \beta$ based on its conditional posterior distribution (Proposition \ref{prop:post}) or generate samples for both $\bm \beta$ and $\tau^2$ according to their conditional posterior distribution (Proposition \ref{prop:post2}).
Following Proposition \ref{prop:pred}, we can generally predict $y(\bm x)$ and confidence intervals conditional on the posterior samples.

In the second approach, we employ EVI-Im solely as an optimization tool for Maximum A Posteriori (MAP), entailing the minimization of $V(\bm x)=-\log\rho^*(\bm x)$. 
This can be done by simply setting the free energy as $\mathcal{F}(\bm x) = V(\bm x)$ and $N=1$.
As a result, the optimization problem \eqref{Min_problem} at the $i$th iteration becomes 
\[
\bm x^{m+1}=\text{arg}\min_{\bm x} \frac{1}{2\tau}||\bm x-\bm x^m||^2-\log V(\bm x),
\]
which is the celebrated proximal point algorithm (PPA) \cite{rockafellar1976monotone}.
Therefore, EVI is a method that connects the posterior sampling and MAP under the same general framework. 
Based on the MAP, we can obtain the posterior mode for $\bm \beta$ (or $\beta$ and $\tau^2$) and the mode of the prediction and the corresponding inference based on the posterior mode. 
For short, we name the first approach EVI-post and the second EVI-MAP. 

\section{Numerical Examples}

In this section, we demonstrate the performances of EVI-GP and compare it with three commonly used GP packages in \verb|R|, which are \verb|gpfit| \citep{macdoanld_gpfit_2019}, \verb|mlegp| \citep{dancik2008mlegp}, \verb|laGP| \citep{gramacy2015local,gramacy2016laGP}. 
The comparison is illustrated via three examples. 
The first one is a 1-dim toy example and the other two are the OTL Circuit and Borehole examples chosen from the online library built by \cite{simulationlib}. 
All three examples are frequently used benchmarks in the computer experiment literature. 
The codes for EVI-GP and all the examples are available on GitHub with the link. \textcolor{blue}{https://github.com/XavierOwen/EVIGP}

The EVI-GP is implemented in \verb|Python|. 
The proximal point optimization in Algorithm \ref{alg:EVI-Im} is solved by the LBFGS function of \verb|Pytorch| library \citep{NEURIPS2019_9015}. 
Some arguments of the EVI-GP codes are set the same for all three examples. 
The argument \verb|history_size| of the LBFGS function is set to 50 and to perform line search, \verb|line_search_fn| is set to \verb|strong_wolfe|, the strong Wolfe conditions.
In the EVI algorithm, the maximum allowed iteration of the inner loop is set to 100 and the maximum allowed iteration of the outer loop (or \emph{maximum epoch}) is set to 500. 
The outer loop of the EVI is terminated either when the maximum epoch is reached or when the convergence condition is achieved. 
We consider the convergence is achieved if $\frac{1}{N}\sum_{i=1}^N||\bm x_i(t)-\bm x_i(t-1)||_2 < \verb|Tol|$ and $\verb|Tol|=1e-8$. 
These parameter settings are done through many experiments and they lead to satisfactory performance in the following examples.

In each example, we use the same pair of training and testing datasets for all the methods under comparison. 
The designs for both datasets are generated via \verb|maximinLHS| procedure from the \verb|lhs| package in \verb|R| \citep{carnell2022lhs}. 
We run 100 simulations. 
In each simulation, we compute the standardized Root Mean Square Prediction Error (RMSPE) on the test data set, which is defined as follows
\[
\text{standardized RMSPE} = \left (\sqrt{\frac{1}{n_{\text{test}}}\sum_i(\hat{y}_i - y_i)^2}\right )\Bigg/ \text{standard deviation of test}(\bm y),
\]
where $\hat{y}_i$ is the predicted value at the test point $\bm x_i$ and $y_i$ is the corresponding true value. 
Box plots of the 100 standardized RMSPEs of all methods are shown for comparison. 

\subsection{One-Dim Toy Example}

In the one-dim example, the data are generated from the test function $y(x)=x\sin(x)+\epsilon$ for $x\in [0,10]$
The size of the training and test data sets are $n=11$ and $m=100$, respectively. 
The variance of the noise is $\sigma^2=0.5^2$. 

\begin{figure}[!htbp]
\centering
\begin{subfigure}{0.45\textwidth}
\centering
\includegraphics[width=\textwidth]{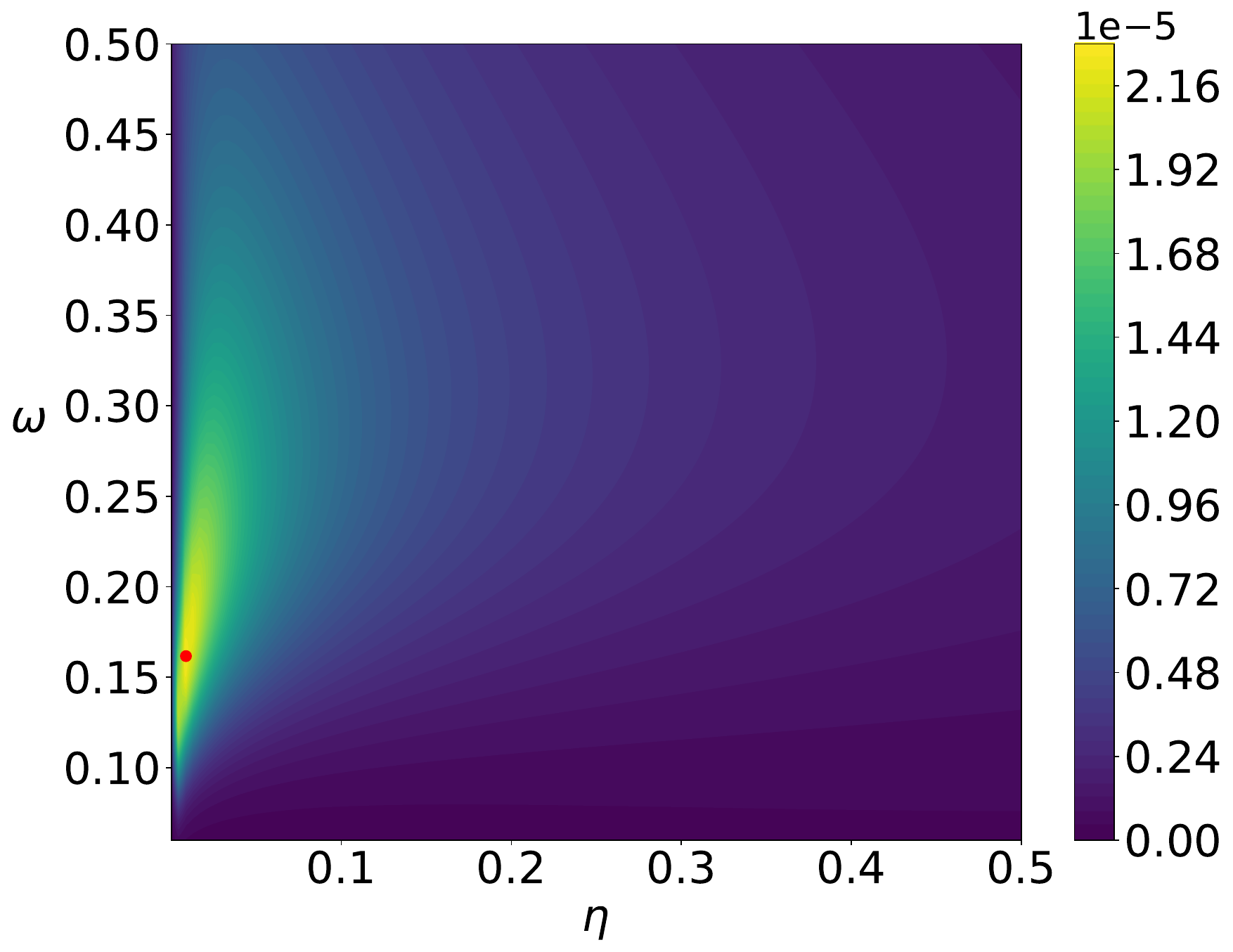}
\caption{GP with constant mean, EVI-MAP.}\label{fig:toy-post-a}
\end{subfigure}
\hfill
\begin{subfigure}{0.45\textwidth}
\centering
\includegraphics[width=\textwidth]{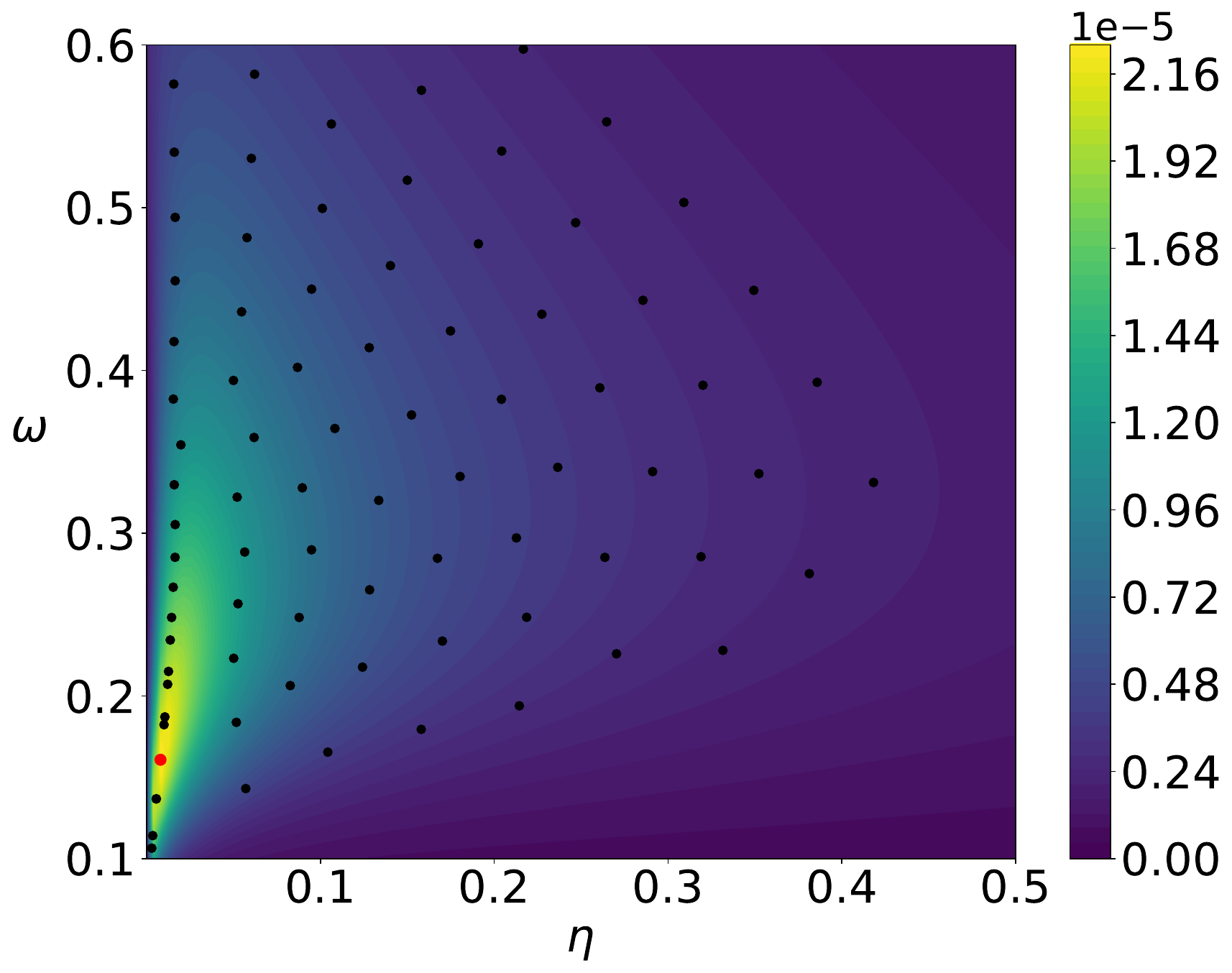}
\caption{GP with constant mean, EVI-post.}\label{fig:toy-post-b}
\end{subfigure}
\hfill
\begin{subfigure}{0.45\textwidth}
\centering
\includegraphics[width=\textwidth]{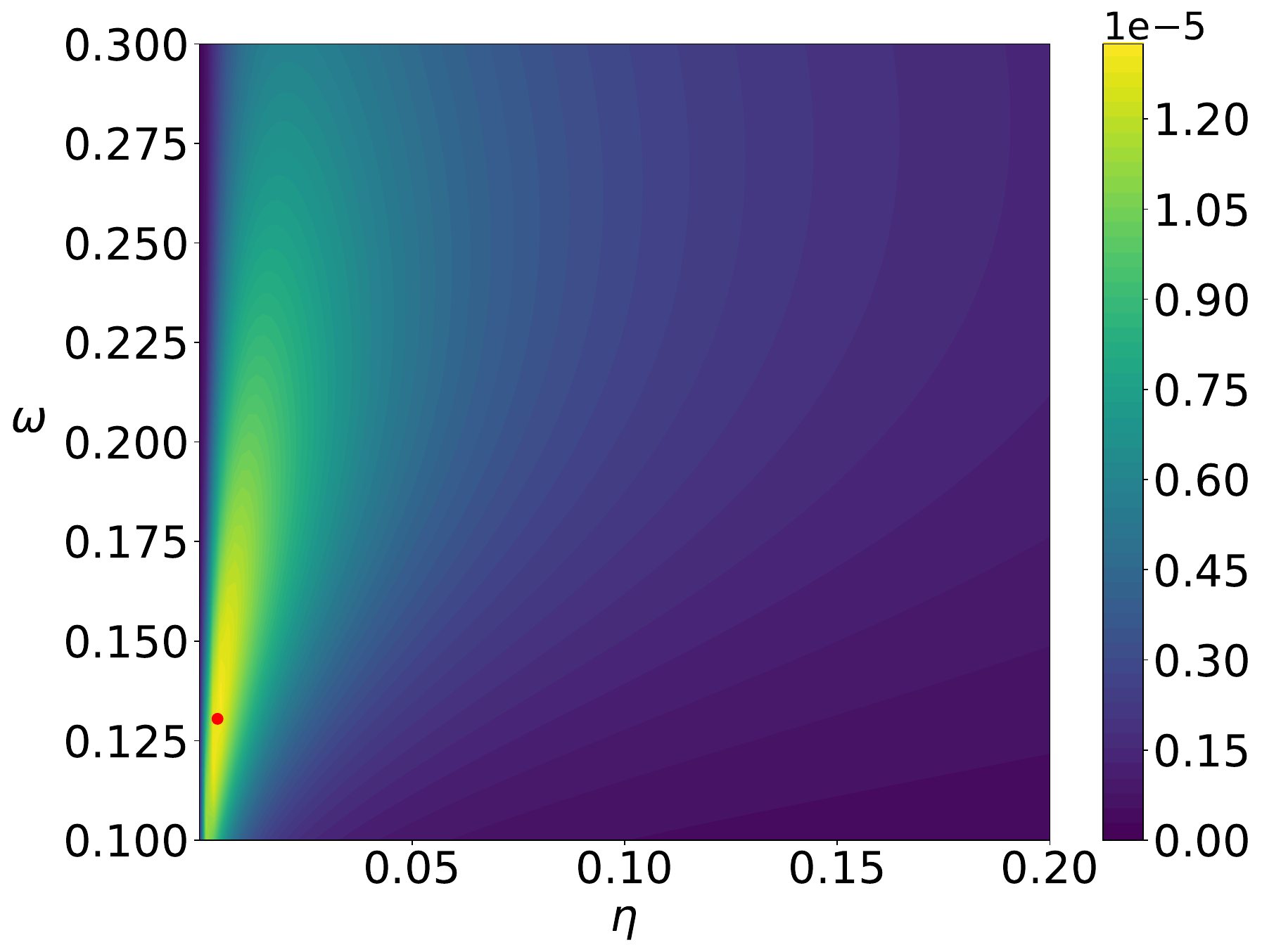}
\caption{GP with linear mean, EVI-MAP.}\label{fig:toy-post-c}
\end{subfigure}
\hfill
\begin{subfigure}{0.45\textwidth}
\centering
\includegraphics[width=\textwidth]{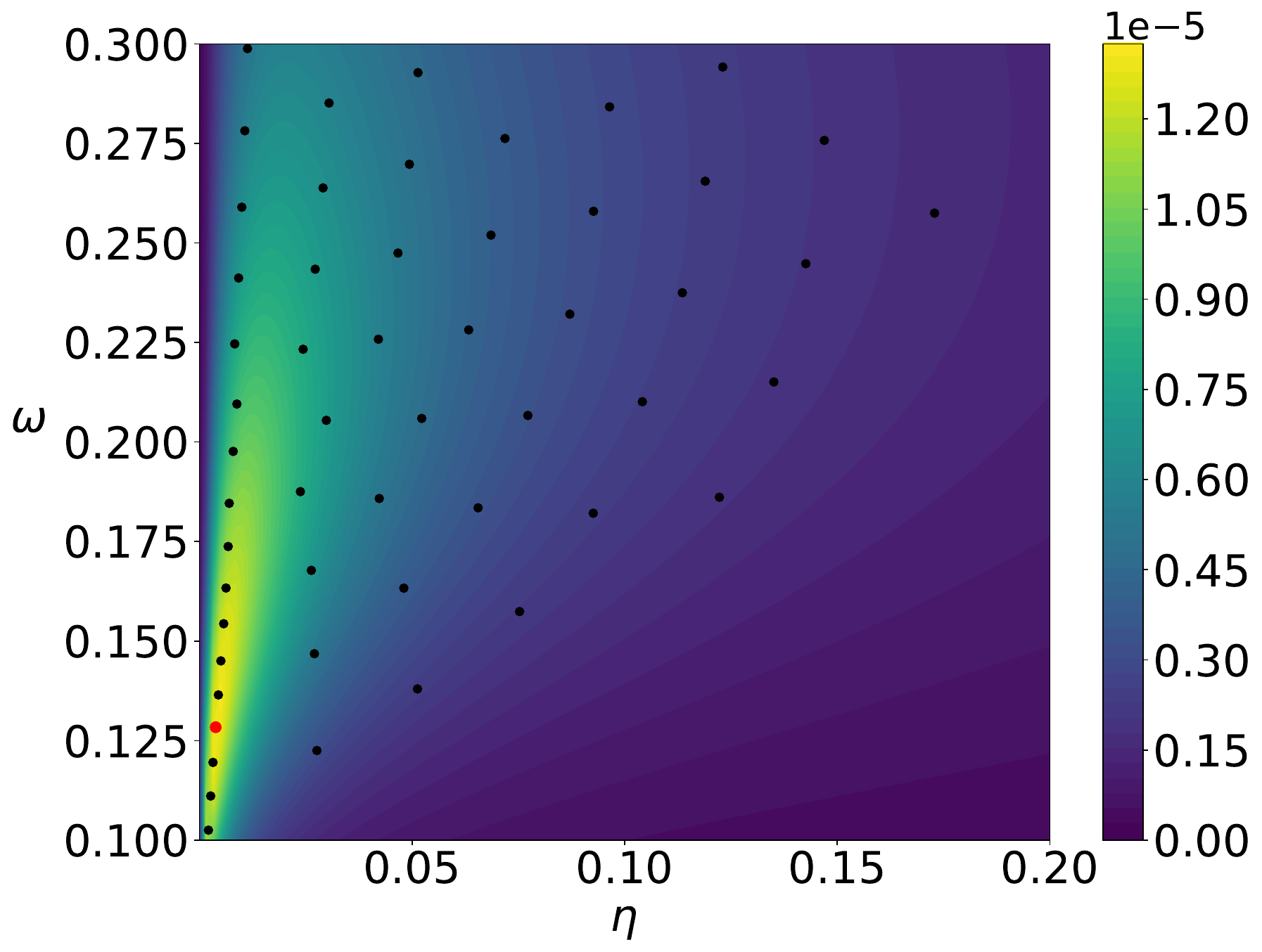}
\caption{GP with linear mean, EVI-post.}\label{fig:toy-post-d}
\end{subfigure}
\caption{One-Dim Toy Example: in Figure \ref{fig:toy-post-a}--\ref{fig:toy-post-d} the contours are plotted according to $p(\omega, \eta)$ evaluated on mesh points without the normalizing constant, the red points are the modes of the posterior distribution of $(\omega, \eta)$ returned by EVI-MAP approach, and the black dots in Figure \ref{fig:toy-post-b} and \ref{fig:toy-post-d} are the particles returned by EVI-post approach.} 
\label{fig:toy-post}
\end{figure}

\begin{figure}[!htbp]
    \centering
 \includegraphics[width=.45\linewidth]{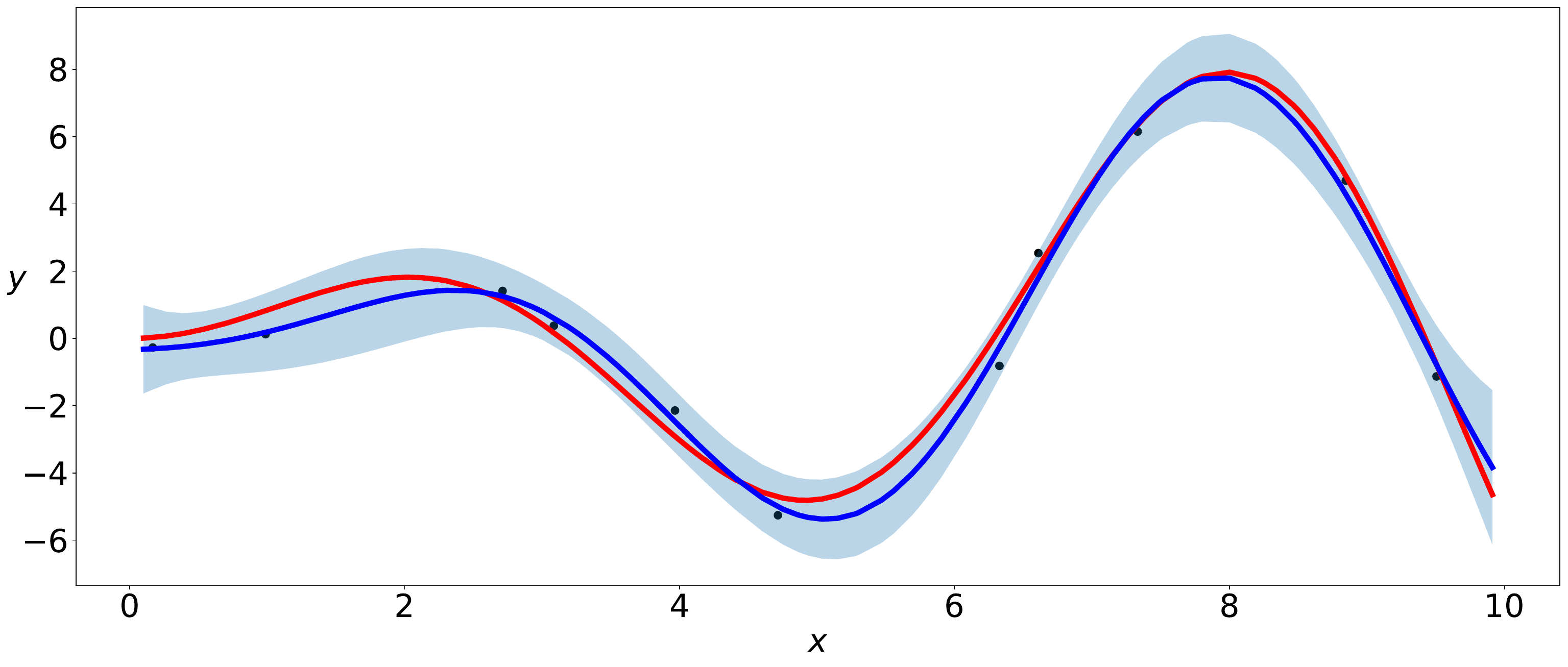}
 \includegraphics[width=.45\linewidth]{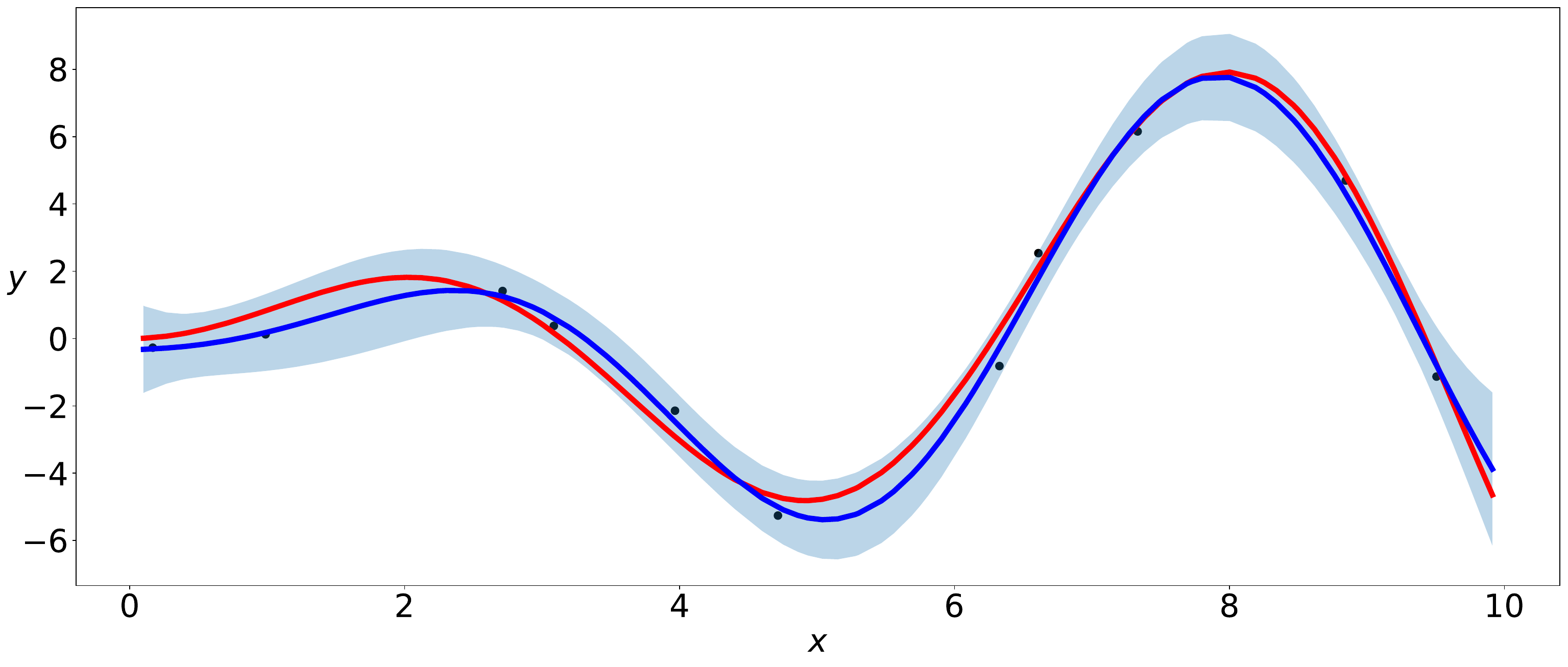}\\
 \includegraphics[width=.45\linewidth]{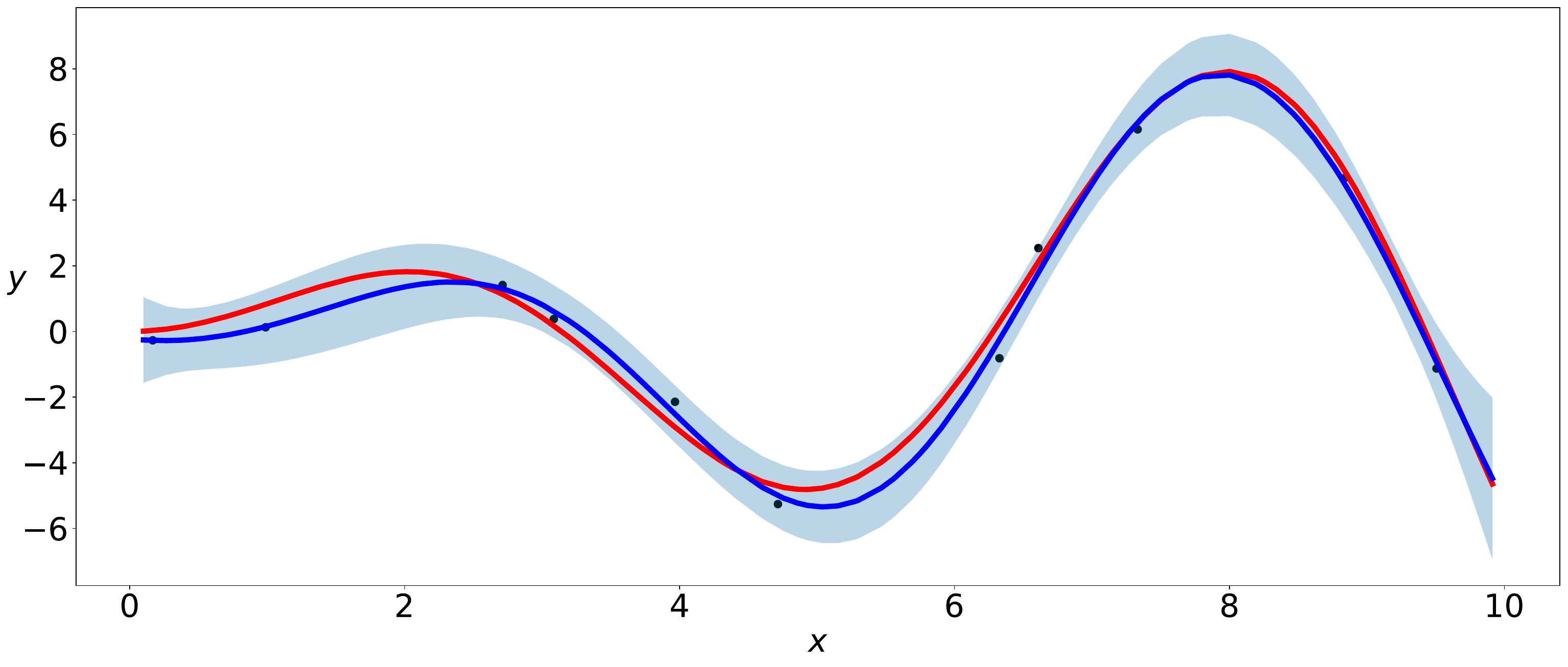}
 \includegraphics[width=.45\linewidth]{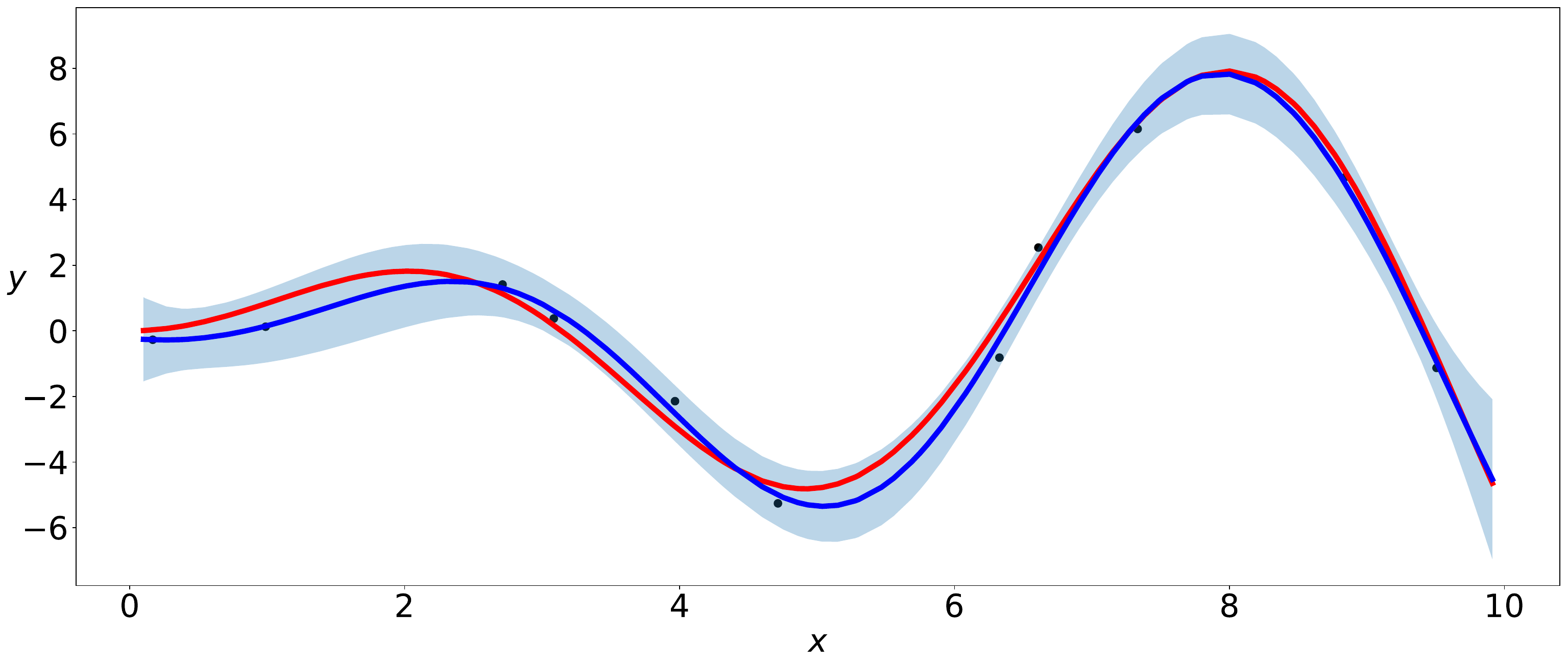}
    \caption{The Gaussian process prediction. Black dots are training data, the red curve is the true \(y(x)=x\sin(x)\), the blue curve is the predicted curve, and the light blue area shows the $95\%$ predictive confidence interval.}
\label{fig:pred-toy}
\end{figure}

The parameters of the prior distributions of the EVI-GP are \(a_\omega = a_\eta=1\), \(b_\omega=b_\eta=0.5\), and \(df_{\tau^2} = 0\) which is equivalent to $p(\tau^2)\propto 1/\tau^2$. 
We consider two possible mean functions for the GP model, the constant mean $\mu(x)=\beta_0$ and the linear mean $\mu(x)=\beta_0+\beta_1x$. 
There is no need for parameter regularization for these simple mean functions, and thus we use non-informative prior for $\bm \beta$. 
We use the EVI-post to approximate $p(\omega, \eta|\bm y_n)$ and set the number of particles $N=100$, kernel bandwidth $h=0.02$ and stepsize $\tau=1$ in the EVI procedure. 
The initial particles are sampled from the uniform distribution in $[0,0.1]\times[0.1,0.4]$. 
Figure \ref{fig:toy-post} and \ref{fig:pred-toy} show the posterior modes and particles returned by EVI-MAP and EVI-post, as well as the prediction and predictive confidence interval returned by both methods. 

\begin{figure}[!ht]
\centering
\includegraphics[width=.99\linewidth]{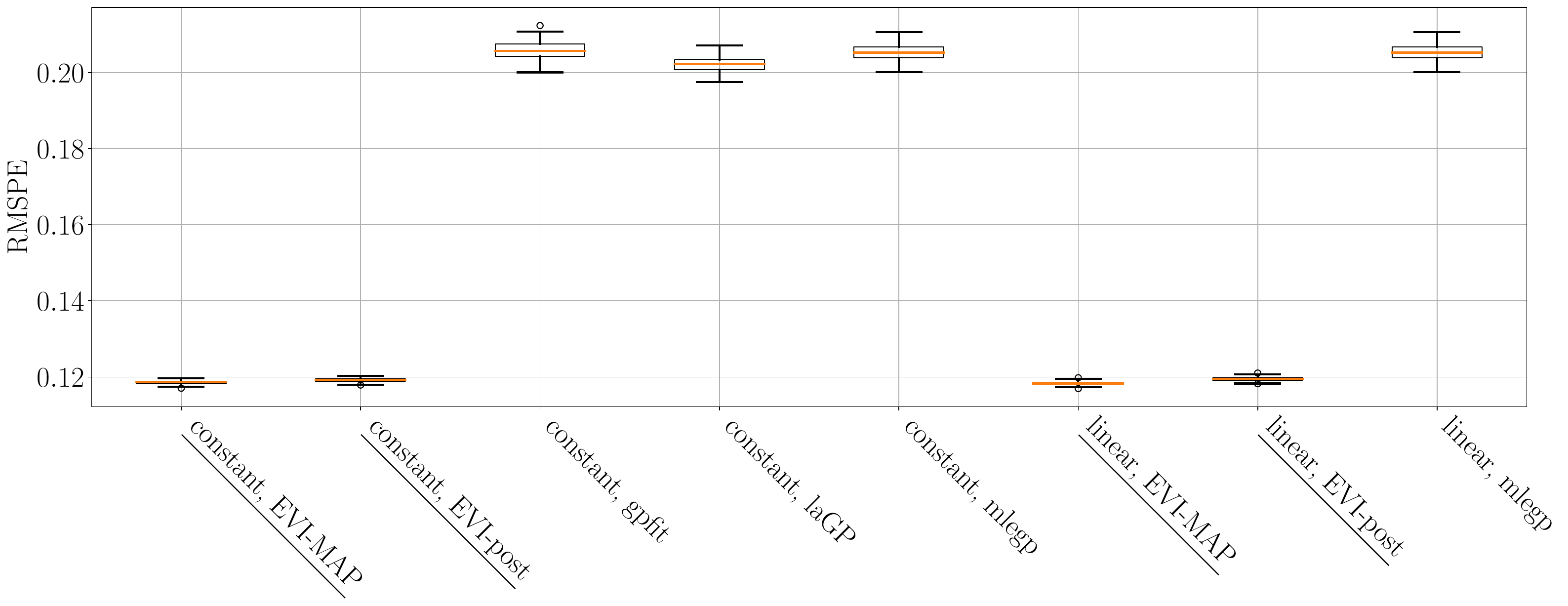}
\caption{Box plots of the standardized RMSPEs of the toy example with all different mean models and different methods. The labels of the proposed EVI-GP methods are underlined.\label{fig:RMSPE-toy}}
\end{figure}

We compare the EVI-GP with three \verb|R| packages.
For the \verb|gpfit| package, we set the nugget threshold to be $[0,25]$, corresponding to $\eta$ in our method, and the correlation is the Gaussian kernel function. 
For the \verb|mlegp| package, we set the argument \verb|constMean| to be 0 for the constant mean model and 1 for the linear mean model. 
The optimization-related arguments in \verb|mlegp| are set as follows. 
The number of simplexes tries is 5. 
Each simplex maximum iteration is 5000. 
The relative tolerance is $1e^{-8}$, BFGS maximum iteration is 500, BFGS tolerance is $0.01$, and BFGS bandwidth is $1e^{-10}$. 
Other parameters in these two packages are set to default. 
For the \verb|laGP| package, we set the argument \verb|start| at 6, \verb|end| at 10, \verb|d| at null, \verb|g| at $1e^{-4}$, \verb|method| to a list of "alc", "alcray", "mspe", "nn", "fish", and \verb|Xi.ret| at True.
The comparison in terms of prediction accuracy is shown in Table \ref{tab:onedim} and Figure \ref{fig:RMSPE-toy}.
Table \ref{tab:onedim} shows the mean of RMSPEs from 100 simulations and Figure \ref{fig:RMSPE-toy} shows the box plots of the RMSPEs. 
In the third column of Table \ref{tab:onedim}, we only list the best result from the three \verb|R| packages. 
The prediction accuracy of both versions of the EVI-GP performs almost equally well and both significantly outperform the three \verb|R| packages. 

\begin{table}[ht]
\centering
\caption{Standardized RMSPE of the One-Dim Toy Example.\label{tab:onedim}}
\begin{tabular}{|c|c|c|c|}\hline
Mean Model & EVI-post & EVI-MAP & gpfit/mlegp/lagp \\\hline
Constant  & 0.1310 & 0.1311 & 0.1500 \\ 
Linear  & 0.1195 & 0.1194 & 0.1584 \\\hline
\end{tabular}
\end{table}

\subsection{OTL-Circuit function}

We test the proposed method using the OTL-circuit function of input dimension $d=6$. 
The OTL-circuit function models an output transformerless push-pull circuit. The function is defined as follows,
\begin{align*}
    V_m(\bm x)& = \frac{(V_{b1}+0.74)I(R_{c2}+9)}{I(R_{c2}+9)+R_f}+\frac{11.35 R_f}{I(R_{c2}+9)+R_f}+\frac{0.74R_f I(R_{c2}+9)}{(I(R_{c2}+9)+R_f)R_{c1}}, 
\end{align*}
where $V_{b1} = 12R_{b2}/(R_{b1}+R_{b2})$, $R_{b1} \in [50, 150]$ is the resistance of b1 (K-Ohms), $R_{b2} \in [25, 70]$ is the resistance of b2 (K-Ohms), $R_f \in [0.5, 3]$ is the resistance of f (K-Ohms), $R_{c1} \in [1.2, 2.5]$ is the resistance of c1 (K-Ohms), $R_{c2} \in [0.25, 1.2]$ is the resistance of c2 (K-Ohms) and $I \in [50, 300]$ is the current gain (Amperes).

The size of training and testing datasets is 200 and 1000, respectively. 
We set the variance of the noise $\sigma^2=0.02^2$. 
For the prior distributions of $\bm \omega$, we let $\omega_i\overset{\text{i.i.d.}}{\sim} \text{Gamma}(a_{\omega}=1, b_{\omega}=2)$ for $i=2,\ldots, 8$, but $\omega_1\sim \text{Gamma}(a_{\omega}=4, b_{\omega}=2)$. 
The prior distribution of $\eta\sim \text{Gamma}(a_{\eta}=1,b_{\eta}=2)$. 
Both non-informative and informative prior distributions are considered. 
When using the informative prior, we set $r=1/3$ and \(df_{\tau^2} = 7\) for the prior distribution of $\bm \beta$ and $\tau^2$. 
For the variance of the prior distribution of $\bm \beta$, we choose $\nu=4.35$, which was the result of 5-fold cross-validation when the biggest mean model is used. 
After variable selection, the 5-fold cross-validation is conducted again to find the optimal $\nu$ to fit the finalized GP model, which leads to $\nu=4.05$. 
In both cross-validation procedures, $\nu$ is searched in an evenly spaced grid from 0 to 5 with $0.05$ grid size.

\begin{figure}[!ht]
    \centering
    \includegraphics[width=.80\linewidth]{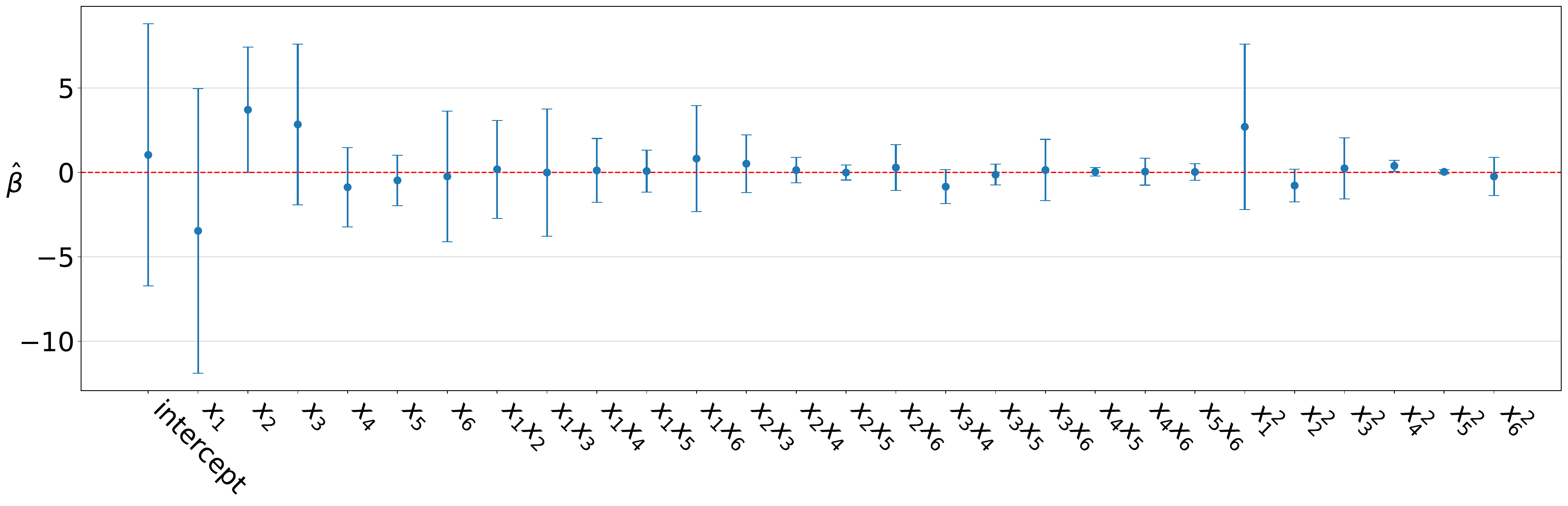}
    \caption{The $95\%$ posterior confidence interval of $\bm \beta$ for the full quadratic mean model.\label{fig:CI-OTL}}
\end{figure}

For both EVI-post and EVI-MAP, we set $h=0.001$ and $\tau = 0.1$. 
For the EVI-post approach, we use $N=100$ particles, and the initial particles of $(\bm \omega, \eta)$ are sampled uniformly in $[0,0.1]^7$. 
The two versions of EVI-GP perform very similarly, so we only return the result of EVI-MAP. 
The initial mean model of the GP is assumed as a quadratic function of the input variables, including the 2-way interactions. 
Based on the $95\%$ posterior confidence interval in Figure \ref{fig:CI-OTL} of the $\bm \beta$, we select $x_2$ and $x_2^2$ as the significant terms to be kept for the final model.

\begin{figure}[!ht]
    \centering
        \includegraphics[width=.90\linewidth]{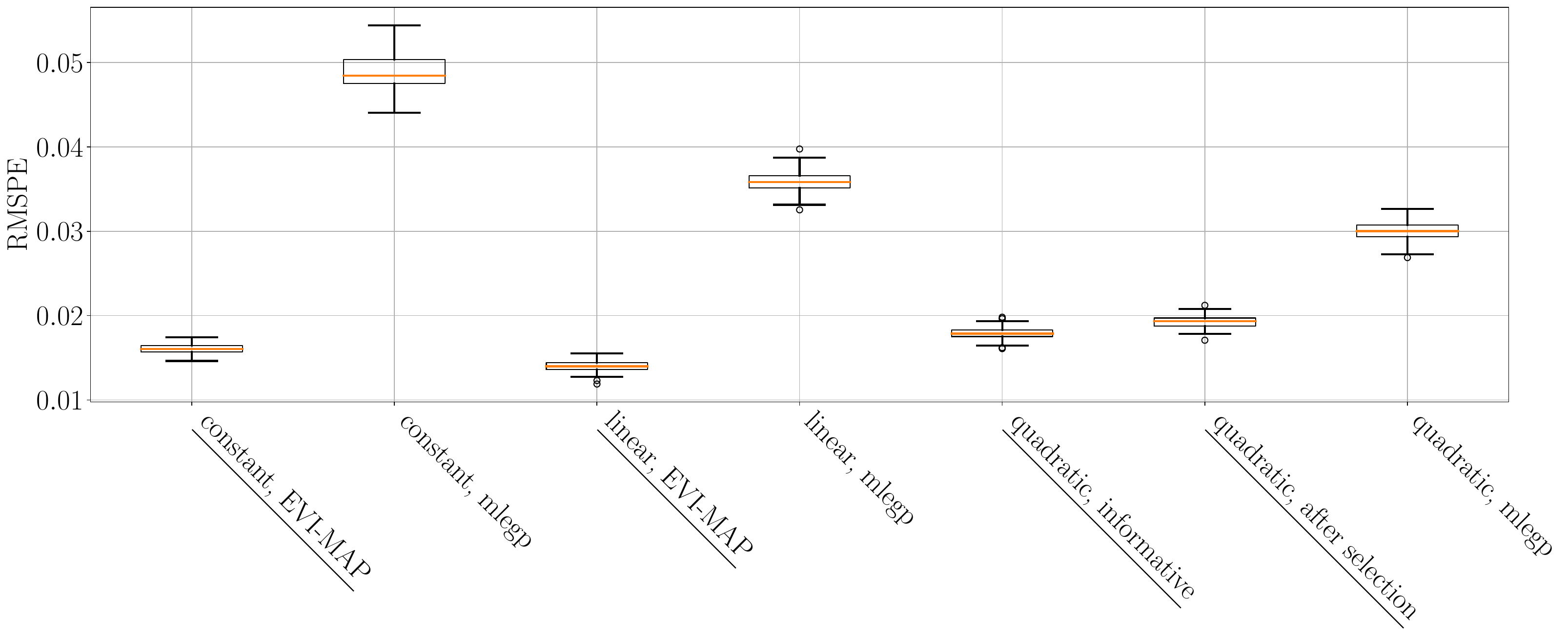}
    \caption{Box plots of standardized RMSPE's of the OTL-Circuit Example.\label{fig:RMSPE-OTL}}
\end{figure}

For comparison, we choose the \verb|R| package \verb|mlegp| because only this package allows the specification of the mean model. 
We set the argument \verb|constMean| to be 0 for the constant mean and 1 for the linear mean model. 
For the full quadratic mean model, we manually add all the second-order effect terms as the input and set \verb|constMean| to 1. 
Unfortunately, the input of the mean and correlation cannot be separately specified in \verb|mlegp|. 
So once we specify the mean part to have all the quadratic terms, the input of the correlation function also contains the quadratic terms. 
But we believe this might give more advantage for \verb|mlegp| because the correlation involves more input variables. 
If some of the variables are not needed, their corresponding correlation parameter estimated by \verb|mlegp| can be close to zero. 
The other arguments in \verb|mlegp| are set in the same way as the previous example. 
Table \ref{tab:OTL} compares the mean of 100 standardized RMSPEs of the 100 simulations of EVI-GP and \verb|mlegp| and Figure \ref{fig:RMSPE-OTL} shows the box plots of the RMSPEs of different models returned by the two methods. 
The EVI-GP is much more accurate than \verb|mlegp| package in terms of prediction. 
However, for this example, variable selection does not improve the model in terms of prediction. 
The most accurate model is the GP with a linear function of the input variables as the mean model.

\begin{table}[ht]
\centering
   \caption{Standardized RMSPE of the OTL-Circuit Example. \label{tab:OTL}}
    \begin{tabular}{ |c|c|c| } 
    \hline
        Type of mean & EVIGP & mlegp \\ \hline
        Constant  & 0.01608 & 0.04882 \\ 
        Linear  & 0.01399  & 0.03584 \\ 
        Quadratic & 0.01792 & 0.03006\\
        Quadratic, after selection & 0.01625 & N/A \\ \hline
    \end{tabular}
 
\end{table}%

\subsection{Borehole function}

In this subsection, we test the EVI-GP with the famous Borehole function, which is defined as follows:
\[
f(\bm x) = \frac{2\pi T_u(H_u-H_l)}{\ln(r/r_\omega) \left(1+\frac{2LT_u}{\ln(r/r_\omega)r_\omega^2 K_\omega}+\frac{T_u}{T_l}\right)},
\]
where $r_\omega \in [0.05, 0.15]$ is the radius of the borehole (m), $r \in [100, 50000]$ is the radius of influence (m), $T_u \in [63070, 115600]$ is the transmissivity of the upper aquifer (m2/yr), $H_u \in [990, 1110]$ is the potentiometric head of upper aquifer (m),$T_l \in [63.1, 116]$ is the transmissivity of the lower aquifer (m2/yr), $H_l \in [700, 820]$ is the potentiometric head of lower aquifer (m), $L \in [1120, 1680]$ is the length of the borehole (m), $K_\omega \in [9855, 12045]$ is the hydraulic conductivity of borehole (m/yr).

\begin{figure}[!ht]
    \centering
    \includegraphics[width=.80\linewidth]{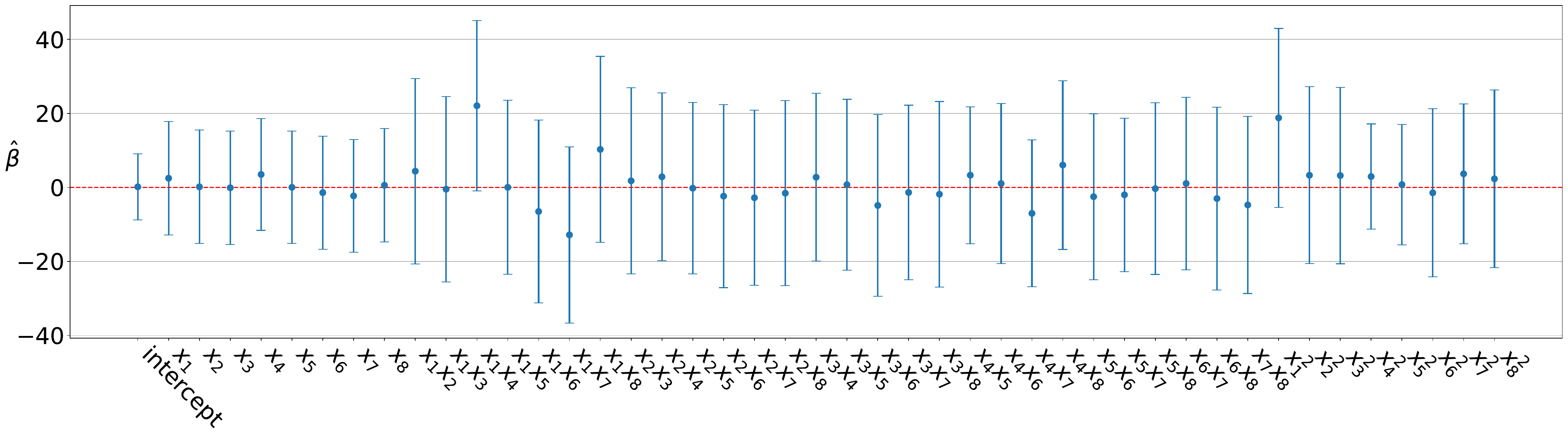}
\caption{The $95\%$ posterior confidence interval of $\bm \beta$ for the full quadratic mean model.\label{fig:CI-Borehole}}
\end{figure}

\begin{figure}[!ht]
\centering
\includegraphics[width=.99\linewidth]{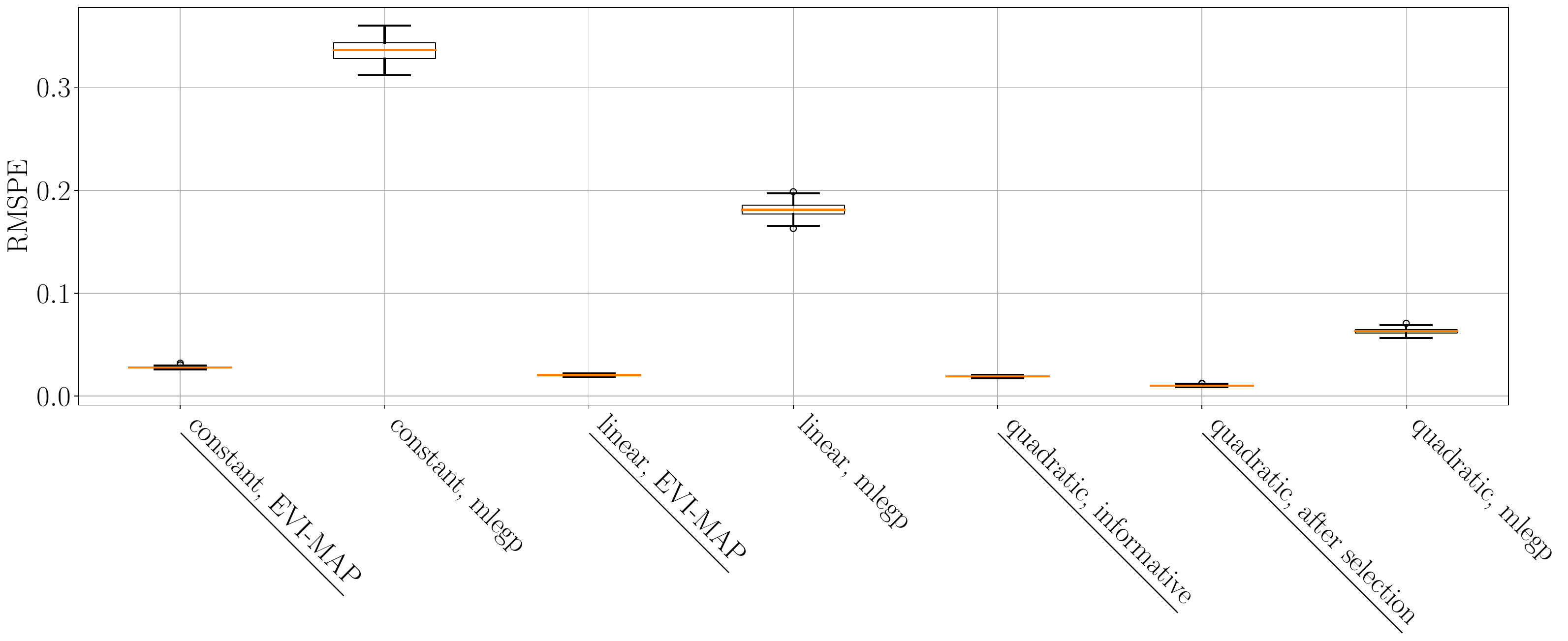}
\caption{Box plots of standardized RMSPE's of the Borehole Example.\label{fig:RMSPE-Borehole}}
\end{figure}

We use a training dataset of size 200 and 100 testing datasets of size 100. 
The noise variance is \(\sigma^2=0.02^2\). 
We use the same parameter settings for the prior distributions and EVI-GP method as in the OTL-Circuit example. 
Regarding the variance of the prior distribution of $\bm \beta$, $\nu=4.55$ was the result of 5-fold cross-validation when the biggest mean model was used. 
After variable selection, the 5-fold cross-validation is conducted again to find the optimal $\nu$ to fit the finalized GP model. 
Coincidently, the optimal $\nu$ is also $4.55$.
The initial mean model of the GP is assumed as a quadratic function of the input variables, including the 2-way interactions. 
Based on the $95\%$ posterior confidence interval in Figure \ref{fig:CI-Borehole} of the $\bm \beta$, we select intercept, $x_1,x_4$, and $x_1x_4$ as the significant terms to be kept for the final model. 
Similarly, we compare the EVI-GP with the \verb|mlegp| package, and the RMSPEs of the 100 simulations are shown in Table \ref{tab:borehole} and Figure \ref{fig:RMSPE-Borehole}. 
Again, EVI-GP is much better than the \verb|R| package. 
More importantly, variable selection proves to be essential for the Borehole example, as the final GP model with the reduced mean model is more accurate than both the quadratic and linear mean models. 

\begin{table}[ht]
\centering
\caption{Standardized RMSPE of the Borehole Example. \label{tab:borehole}}
\begin{tabular}{ |c|c|c| } 
\hline
Type of mean & EVIGP & mlegp \\\hline
Constant  & 0.01151 & 0.3354 \\ 
Linear  & 0.04191  & 0.1810 \\ 
Quadratic & 0.01212 & 0.2019\\
Quadratic, after selection & 0.01019 & N/A\\ \hline
\end{tabular}    
\end{table}%

\section{Conclusion}

In this book chapter, we review the conventional Gaussian process regression model under the Bayesian framework. 
More importantly, we propose a new variational inference approach, called Energetic Variational Inference, as an alternative to traditional MCMC approaches to estimate and make inferences for the GP regression model. 
Through comparing with some commonly used \verb|R| packages, the new EVI-GP performs better in terms of prediction accuracy. 
Although not completely revealed in this book chapter, the true potential of variational inference lies in transforming the MCMC sampling problem into an optimization problem. 
As a result, it can be used to solve a more complicated Bayesian framework. 
For instance, we can adapt the GP regression and classification by adding fairness constraints on the parameters such that it can meet the ethical requirements in many social and economic contexts. 
In this case, variational inference can easily solve the constrained optimization whereas it is very challenging to do MCMC sampling with fairness constraints. 
We are pursuing in this direction in the future work. 

\section*{Acknowledgment}

The authors' work is partially supported by NSF Grant \# 2153029 and \# 1916467. 

\section*{Appendix}

\noindent{\bf Derivation of Proposition \ref{prop:post}}
\begin{proof}
The marginal posterior distribution of $(\bm \omega, \tau^2, \eta)$ can be obtained by integration as follows. 
\begin{align*}
& p(\bm \omega, \tau^2,\eta|\bm y_n) \propto \int p(\bm \beta) p(\bm y_n|\bm \theta) p(\bm \omega)p(\tau^2)p(\eta)\dd \bm \beta\\
\propto & \int\left\{ \exp\left[-\frac{1}{2}(\bm \beta-\hat{\bm \beta}_n)^\top \bm \Sigma_{\bm \beta|n}^{-1}(\bm \beta-\hat{\bm \beta}_{n})\right]\det(\bm \Sigma_{\bm \beta|n})^{-1/2} \right.\\
 & \det(\bm \Sigma_{\bm \beta|n})^{1/2}\exp\left[-\frac{1}{2}\hat{\bm \beta}_n^\top \bm \Sigma_{\bm \beta|n}^{-1}\hat{\bm \beta}_n-\frac{1}{2\tau^2}\bm y_n^\top (\bm K_n+\eta \bm I_n)^{-1}\bm y_n\right]\\
& \left.(\tau^2)^{-n/2}\det(\bm K_n+\eta \bm I_n)^{-1/2}p(\tau^2)p(\bm \omega)p(\eta)\right\} \dd \bm \beta \\
\propto & \det(\bm \Sigma_{\bm \beta|n})^{1/2}\exp\left[-\frac{1}{2}\hat{\bm \beta}_n^\top \bm \Sigma_{\bm \beta|n}^{-1}\hat{\bm \beta}_n-\frac{1}{2\tau^2}\bm y_n^\top (\bm K_n+\eta \bm I_n)^{-1}\bm y_n\right]\\
&\times (\tau^2)^{-n/2}\det(\bm K_n+\eta \bm I_n)^{-1/2}p(\tau^2)p(\bm \omega)p(\eta). 
\end{align*}
\end{proof}

\noindent{\bf Derivation of Proposition \ref{prop:post2}}

Based on the non-informative prior, 
\begin{align*}
\bm \Sigma_{\bm \beta|n}&=\tau^2\left[\bm G^\top (\bm K_n+\eta \bm I_n)^{-1} \bm G\right]^{-1}\\
\hat{\bm \beta}_n &= \left[\bm G^\top (\bm K_n+\eta \bm I_n)^{-1} \bm G\right]^{-1}\left[\bm G^\top (\bm K_n+\eta \bm I_n)^{-1}\right]\bm y_n.
\end{align*}
Define 
\begin{align*}
s^2_n&=\tau^{-2} \hat{\bm \beta}_n^\top \bm \Sigma_{\bm \beta | n}^{-1}\hat{\bm \beta}_n+\bm y_n^\top (\bm K_n+\eta \bm I_n)^{-1}\bm y_n\\
&=\bm y_n^\top \left[(\bm K_n+\eta \bm I_n)^{-1}\bm G \left(\bm G^\top (\bm K_n+\eta \bm I_n)^{-1}\bm G\right)^{-1}\bm G^\top (\bm K_n+\eta \bm I_n)^{-1} + (\bm K_n+\eta \bm I_n)^{-1}\right]\bm y_n\\
&=\bm y_n^\top (\bm K_n+\eta \bm I_n)^{-1}\left[\bm G \left(\bm G^\top (\bm K_n+\eta \bm I_n)^{-1}\bm G\right)^{-1}\bm G^\top + (\bm K_n+\eta \bm I_n)\right](\bm K_n+\eta \bm I_n)^{-1}\bm y_n.
\end{align*}
Then $p(\bm \omega, \tau^2,\eta |\bm y_n)$ is 
\begin{align*}
& p(\bm \omega, \tau^2,\eta |\bm y_n) \propto (\tau^2)^{p/2}\exp\left(-\frac{s_n^2}{2\tau^2}\right)(\tau^2)^{-n/2} p(\tau^2) \det(\bm G^\top (\bm K_n+\eta \bm I_n)^{-1}\bm G)^{-1/2}\det(\bm K_n+\eta \bm I_n)^{-1/2}p(\bm \omega) p(\eta)\\
& \propto (\tau^2)^{-\left[\frac{1}{2}(df_{\tau^2}+n-p)+1\right]}\exp\left(-\frac{1+s_n^2}{2\tau^2}\right)\det(\bm G^\top (\bm K_n+\eta \bm I_n)^{-1}\bm G)^{-1/2}\det(\bm K_n+\eta \bm I_n)^{-1/2}p(\bm \omega) p(\eta).
\end{align*}
Due to the conditional conjugacy, we can see the conditional posterior distribution of $\tau^2| \bm \omega, \eta, \bm y_n$ is 
\begin{equation}
\tau^2 |\bm \omega, \eta, \bm y_n \sim \text{Scaled Inverse-}\chi^2(df_{\tau^2}+n-p, \hat{\tau}^2), 
\end{equation}
where 
\[
\hat{\tau}^2=\frac{1+s_n^2}{df_{\tau^2}+n-p}.
\]
Based on it, we can obtain the integration
\begin{align*}
p(\bm \omega,\eta |\bm y_n) \propto (\hat{\tau}^2)^{-\frac{1}{2}(df_{\tau^2}+n-p)}\det(\bm G^\top (\bm K_n+\eta \bm I_n)^{-1}\bm G)^{-1/2}\det(\bm K_n+\eta \bm I_n)^{-1/2}p(\bm \omega) p(\eta).
\end{align*}

\bibliographystyle{asa}
\bibliography{EVIGP.bib}

\end{document}